\documentclass[twocolumn,amsmath,amssymb,floatfix]{revtex4}

\usepackage{latexsym,amsmath,amssymb,theorem,epsfig}
\usepackage{graphicx}

\usepackage{bm}
\usepackage{amsmath}
\def\lsim{\mathrel{\mathstrut\smash{\ooalign{\raise2.5pt\hbox{$<$}\cr\lower2.5pt\hbox{$\sim$}}}}}
\def\gsim{\mathrel{\mathstrut\smash{\ooalign{\raise2.5pt\hbox{$>$}\cr\lower2.5pt\hbox{$\sim$}}}}}

\def\myputfigure#1#2#3#4#5%
{\vskip#5pt\makebox[0pt]{\hskip#2in
\includegraphics[width=#3\textwidth]{#1}}\vskip#4pt\hfill}

\def\be{\begin{equation}}
\def\ee{\end{equation}}
\def\bea{\begin{eqnarray}}
\def\eea{\end{eqnarray}}

\begin{document} 

\title{Non-Gaussianities in New Ekpyrotic Cosmology}

\author{Evgeny I. Buchbinder$^{1}$, Justin Khoury$^{1}$, Burt A. Ovrut$^{2}$}

\affiliation{$^1$Perimeter Institute for Theoretical Physics, 
31 Caroline St. N., Waterloo, ON, N2L 2Y5, Canada\\
$^{2}$Department of Physics, The University of Pennsylvania, Philadelphia, PA 19104--6395, USA} 

\begin{abstract}
New Ekpyrotic Cosmology is an alternative scenario of the early universe which relies on a phase of slow contraction before the big bang.
We calculate the 3-point and 4-point correlation functions of primordial density perturbations and
find a generically large non-Gaussian signal, just below the current sensitivity level of CMB experiments. This is in contrast with slow-roll inflation,
which predicts negligible non-Gaussianity. 
The model is also distinguishable from alternative inflationary scenarios that can yield large
non-Gaussianity, such as DBI inflation and the simplest curvaton-like models, through the shape dependence of the correlation functions. 
Non-Gaussianity therefore provides a distinguishing and testable prediction of New Ekpyrotic Cosmology. 
\end{abstract}

\maketitle

The ekpyrotic scenario is a candidate theory of early-universe cosmology. Instead of invoking a short burst of accelerated expansion from a hot initial state,
as in inflation, 
the
ekpyrotic scenario
relies on a cold beginning followed by a phase of very slow contraction. Despite such diametrically opposite dynamics, both models predict a flat, 
homogeneous and isotropic universe, endowed with a nearly scale invariant spectrum of density perturbations, and are, therefore, equally successful at accounting
for all current cosmological observations.

An important drawback of the original ekpyrotic theory~\cite{oldek} is how to avoid the big crunch singularity without introducing ghosts or other pathologies. Moreover, the fate of perturbations
through the bounce is ambiguous --- whereas the scalar field fluctuations are scale invariant during the contracting phase, the curvature perturbation on uniform-density hypersurfaces, $\zeta$, is not. And since the latter remains constant on super-horizon scales, one generally expects its (unacceptably blue) spectrum to be preserved irrespective of the bounce physics. 
Despite considerable work indicating that stringy effects at the bounce may positively alter this conclusion~\cite{tolley}, the issue of matching conditions remains controversial.

Both of these issues have been resolved in the recently proposed New Ekpyrotic scenario~\cite{newek}. In~\cite{newek}, we derived a fully {\it non-singular} bounce within a controlled
and ghost-free four-dimensional effective theory using the ghost condensation mechanism~\cite{ghost}. Moreover, $\zeta$ acquires a scale invariant spectrum well
before the bounce, thanks to an entropy perturbation generated by a second scalar field~\cite{newek,lehners,paolo}. Thus New Ekpyrotic Cosmology appears to be a consistent alternative to the inflationary scenario.

A distinguishing prediction lies in the tensor spectrum~\cite{oldek}: inflation predicts scale invariant primordial gravity waves, whereas ekpyrosis does not.
Detecting tensor modes from CMB B-mode polarization could rule out the ekpyrotic scenario, whereas an absence of detection would not discriminate between the two models.

In this Letter we focus on another key observable: the non-Gaussianity of primordial density perturbations. We show that New Ekpyrotic Cosmology generically predicts a large level of non-Gaussianity, potentially just below current sensitivity levels and detectable by near-future CMB experiments. 

We calculate the 3-point and 4-point functions.  For typical parameter values, the amplitude of the 3-point function is generically large, with $f_{\rm NL}$
around the current WMAP bound~\cite{WMAP}: $-36 < f_{\rm NL} < 100$. That is, assuming all parameters are ${\cal{O}}(1)$, $f_{\rm NL}$ approaches the limits of this bound, depending on the sign of a parameter. These values are well above the expected sensitivity of the Planck experiment: $|f_{\rm NL}|\lsim 20$. 
The amplitude of the 4-point function is also generically large: $\tau_{\rm NL}\sim 10^4$, which is again near the estimated WMAP bound and within the reach of Planck: $\tau_{\rm NL}\lsim 600$~\cite{tauwmap}.

This is in stark contrast with the highly Gaussian spectrum predicted by slow-roll inflation. 
Comparably large non-Gaussianity does arise in non-slow roll models, such as DBI inflation~\cite{DBI}, and whenever the precursor of density fluctuations is a
light spectator field, such as in the curvaton~\cite{curvaton,lythng} or modulon scenarios~\cite{dgz,zal}. 
However, as we will see, New Ekpyrosis predicts a different shape dependence in momentum space for the 3-~and/or 4-point spectrum than the simplest 
such models.

Non-Gaussianity therefore offers a distinguishing prediction of New Ekpyrotic Cosmology, potentially testable in CMB experiments within the next few years.
 
\section{New Ekpyrotic Cosmology}

As with inflation, ekpyrosis relies on a scalar field $\phi$ rolling down a potential ${\cal V}(\phi)$. Instead of being flat and positive, 
however, here ${\cal V}(\phi)$ must be steep, negative and nearly exponential in form. For concreteness, we take
\be
{\cal V}(\phi) = -V_0e^{-\phi/\Lambda}\,,
\label{Vphi}
\ee
where $\Lambda \equiv \sqrt{\epsilon}M_{\rm Pl}$ and $\epsilon \ll 1$. The Friedmann and scalar field equations then yield a background scaling solution,
\be
a(t)\sim (-t)^{2\epsilon}\,;\;\; \bar{\phi}(t)= \Lambda\log\left(\frac{V_0}{2\Lambda^2(1-6\epsilon)} \; t^2\right)\,,
\label{scaling}
\ee
with Hubble parameter $H=2\epsilon/t$.
Since $\epsilon\ll 1$, this describes a slowly-contracting universe with rapidly increasing $H$, again in contrast with the rapid expansion and nearly constant $H$ in inflation.

In single-field ekpyrosis, fluctuations in $\phi$ acquire a scale invariant spectrum. As we review shortly, this traces back to the fact that the above solution
satisfies $\bar{{\cal V}}_{,\phi\phi} = -2/t^2$. However this contribution exactly projects out of $\zeta$, leaving the latter with an unacceptably blue spectrum.
Since $\zeta$ is conserved on super-horizon scales barring entropy perturbations, it is generally expected to match continuously through the bounce,
although stringy effects could alter this picture~\cite{tolley}.

New Ekpyrotic Cosmology introduces a second field, $\chi$, as the progenitor of the scale-invariant perturbation spectrum~\cite{newek,lehners}.
This field has no dynamics during the ekpyrotic phase and remains approximately fixed at 
$\bar{\chi}= 0$. However, as we describe below, its fluctuations generate a scale-invariant spectrum of entropy perturbations, which gets imprinted onto $\zeta$ at the end of the ekpyrotic phase. 

An essential condition in obtaining a scale-invariant spectrum is that at $\bar{\chi}=0$ the curvature of the potential be nearly the same along the $\chi$ and $\phi$ directions:
$\bar{V}_{,\chi\chi} \approx \bar{V}_{,\phi\phi}$. An example of such a potential is
\be
V(\phi,\chi) ={\cal V}(\phi) \left(1+\frac{\chi^2}{2\Lambda^2}  + \frac{\alpha_3}{3!} \frac{\chi^3}{\Lambda^3} +  \frac{\alpha_4}{4!} \frac{\chi^4}{\Lambda^4} + \ldots \right)\,.
\label{V2field}
\ee
The higher-order $\chi$ terms are naturally expected to be
suppressed by the same scale $\Lambda$ as the quadratic term, hence the form~(\ref{V2field}). For simplicity we take $\alpha_3,\alpha_4,\ldots$ to be constants.
While potential~(\ref{V2field}) yields a slightly blue spectral tilt, a more general potential
is presented in~\cite{newek} which allows for the observed red tilt without altering
the conclusions for non-Gaussianity arrived at in this paper.
Note that the required field trajectory lies along an unstable point. However, a pre-ekpyrotic,
stabilizing phase can easily create initial conditions so that this trajectory is arbitrarily close to the tachyonic ridge~\cite{newek}.

{\it Power spectrum for $\chi$}: Since our space-time background is nearly static, we ignore gravity in studying $\chi$ perturbations. To linear order, the Fourier modes $\delta\chi_k^{(0)}$ around 
$\bar{\chi}=0$ satisfy a free field equation with time-dependent mass $\bar{V}_{,\chi\chi} = \bar{V}_{,\phi\phi} = -2/t^2$:
\be
\ddot{\delta \chi_k}^{(0)} + \left(k^2 -\frac{2}{t^2}\right)\delta\chi_k ^{(0)}= 0 \,.
\label{dchi0}
\ee
Assuming the usual adiabatic vacuum, we find
\be
\delta\chi_k^{(0)} = \frac{e^{-ikt}}{\sqrt{2k}}\left(1-\frac{i}{kt}\right)\,.
\label{freedchi}
\ee
On super-Hubble scales, $k(-t)\ll 1$, the power spectrum, defined by $\langle \delta \chi_k^{(0)}
\delta \chi_{k'}^{(0)}\rangle = (2\pi)^3\delta^3(\vec{k}+\vec{k}')P_\chi(k)$, is
\be
k^3P_\chi(k) = \frac{1}{2t^2}\,,
\label{burt}
\ee
which is scale invariant. Including gravity and departing from the pure exponential form~(\ref{Vphi}) results in small deviations from scale invariance.
This can yield a small red tilt, consistent with current CMB observations~\cite{newek}. 

{\it Evolution of $\zeta$}: We focus for simplicity on the regime where all relevant modes are well-outside the horizon, $k \ll aH$. In the small-gradient approximation,
the metric can be written as $ds^2 = -{\cal N}^2dt^2 + e^{2\zeta(\vec{x},t)}a^2(t)d\vec{x}^2$~\cite{gradient}, where ${\cal N}$ is the lapse function, and $\zeta$ is the curvature perturbation. The evolution of $\zeta$ on uniform-density hypersurfaces is governed by
\be
\dot{\zeta} = 2H\frac{\delta V}{\dot{\bar{\phi}}^2 - 2\delta V} \,,
\label{dotzeta}
\ee
where $\delta V\equiv V(\phi,\chi) - V(\bar{\phi},\bar{\chi})$.
A key simplification is that $\delta\phi$ has a steep blue spectrum at long wavelengths
and, hence, can be neglected. Thus, for the potential~(\ref{V2field}), we have $\delta V \approx {\cal V}(\bar{\phi})\delta\chi^2/2\Lambda^2 + \ldots$

To proceed further, one needs an expression for $\delta\chi$ to higher-order than the ``free" part $\delta\chi^{(0)}$. 
To do this, we solve $\ddot{\delta\chi} + \bar{V}_{\chi\chi}\delta\chi = 0$, valid at long wavelengths, perturbatively: 
$\delta\chi = \delta\chi^{(0)} + \delta\chi^{(1)}+\ldots$ To lowest order, this equation reduces to~(\ref{dchi0}) in the limit $k\rightarrow 0$. 
The next order, $\delta\chi^{(1)}$, satisfies $\ddot{\delta\chi}^{(1)} + \bar{V}_{\chi\chi}\delta\chi^{(1)} + \bar{V}_{,\chi\chi\chi} (\delta\chi^{(0)})^2/2 = 0$.
Using~(\ref{scaling}),~(\ref{V2field}) and $\delta\chi^{(0)}\sim 1/t$, we find
\be
\delta\chi = \delta\chi^{(0)} + \frac{\alpha_3}{4\Lambda} \left(\delta\chi^{(0)}\right)^2+\ldots
\label{pertdchi}
\ee
Substituting into~(\ref{dotzeta}), one can integrate to obtain
\be
\zeta_{\rm ek} = \frac{1}{2}\left(\frac{\delta\chi^{(0)}}{M_{\rm Pl}}\right)^2 + \frac{5\alpha_3}{18\sqrt{\epsilon}}\left(\frac{\delta\chi^{(0)}}{M_{\rm Pl}}\right)^3 + \ldots 
\label{zetaek}
\ee

The ekpyrotic phase must eventually end if the universe is to undergo a smooth bounce and reheat into a hot big bang phase.
This is achieved by adding a feature to the potential~(\ref{V2field}) which eventually pushes $\chi$ away from the tachyonic ridge~\cite{newek}.
Denote the time at which ekpyrosis stops as $t_{\rm end}$.
For simplicity, we model this with $V_{,\chi}$ suddenly becoming non-zero and nearly constant at $\chi=0$. Denote this constant by $V_{,\chi}|$. The exit phase is assumed to last
for a time interval $\Delta t$ which is short compared to a Hubble time: $|H_{\rm end}|\Delta t \ll 1$. This will be the case provided
the potential satisfies
\be
\epsilon_\chi \equiv \frac{H^4_{\rm end}M_{\rm Pl}^2}{V_{,\chi}|^2} \lsim 1\,. 
\ee

The exit phase generates an additional contribution to $\zeta$. To compute this
in the rapid-exit approximation, we can treat the right-hand side of~(\ref{dotzeta}) as approximately constant. In  
evaluating this constant, note that, to leading order in $\delta\chi$, we have $\delta V \approx V_{,\chi}|\delta\chi = \pm H^2_{\rm end}M_{\rm Pl}\delta\chi/\sqrt{\epsilon_\chi}$. (Higher-order terms in $\delta\chi$ yield small corrections to~(\ref{zetaek}) and are therefore negligible.) 
Thus $\zeta$ changes from $\zeta_{\rm ek}$ by an amount $\zeta_c$ during the exit, given by
\be
\zeta_c = \mp 2\sqrt{\epsilon}\beta\frac{\delta\chi(t_{\rm end})}{M_{\rm Pl}}\,,
\label{zetac}
\ee
where $\beta\equiv |H_{\rm end}|\Delta t \sqrt{\epsilon/\epsilon_\chi}$. Noting that to lowest order $\delta\chi\approx\delta\chi^{(0)}$ and substituting~(\ref{burt}) evaluated at $t_{\rm end}$, it follows
from~(\ref{zetac}) that the $\zeta$ power spectrum is
\be
k^3P_{\zeta}(k)=\frac{4\epsilon\beta^2}{M_{\rm Pl}^2} k^3 P_{\chi}(k) = \beta^2\frac{H_{\rm end}^2}{2\epsilon M_{\rm Pl}^2}\,.
\label{burt2}
\ee
Up to the prefactor $\beta^2$, this is identical to the inflationary result, with $\epsilon$ playing the role of the usual slow-roll parameter. 
In the exit mechanism of~\cite{newek}, $\beta$ denotes the overall change in angle in the field trajectory: $\beta = \Delta\theta$.

Let us pause to discuss the parameter values that satisfy the CMB constraint $k^3P_{\zeta}(k) \approx 10^{-10}$.
Although $H$ passes through zero at the bounce, as argued in~\cite{newek} 
its magnitude is essentially the same at the beginning of the hot big bang phase as it was at 
$t_{\rm end}$, the end of the ekpyrotic phase. In other words,
$H_{\rm end}$ sets the reheat temperature in the expanding phase. For GUT-scale reheat temperature, we have $H_{\rm end}/M_{\rm Pl}\approx 10^{-6}$. 
Meanwhile, $\beta$ is a free parameter whose value depends on the exit dynamics. For the explicit exit mechanism of~\cite{newek}, however, the natural value is
$\beta\sim {\cal O}(1)$. In this case, setting $k^3P_{\zeta}(k) = 10^{-10}$ implies $\epsilon \approx 10^{-2}$. 
We will henceforth take $\beta = 1$ and $\epsilon =  10^{-2}$ as fiducial parameter values.

Combining~(\ref{zetac}) with~(\ref{pertdchi}) and ~(\ref{zetaek}) yields
\be
\zeta(x) = \zeta_c(x) +\frac{1}{8\epsilon\beta^2}\zeta_c^2(x)\mp\frac{5\alpha_3}{144\epsilon^2\beta^3}\zeta^3_c(x) + \ldots \,.
\label{zetanonlin}
\ee
The exit from the ekpyrotic phase is followed by a ghost condensate phase which leads
to a {\it non-singular} bounce and reheating. Meanwhile, $\chi$ gets stabilized and further evolution is governed by the single scalar $\phi$.
It follows that $\zeta$ is conserved through the bounce and emerges unscathed in the hot big bang phase.

\section{Non-Gaussianity}

{\it 3-point function}: The 3-point $\zeta$ correlation function in 
New Ekpyrotic Cosmology is given by~\cite{newek}
\be
\langle \zeta_{k_1} \zeta_{k_2} \zeta_{k_3}\rangle = (2\pi)^3\delta^3(\vec{k}_1+\vec{k}_2+\vec{k}_3)B(k_1,k_2,k_3)\,,
\label{14}
\ee
where the shape function $B(k_1,k_2,k_3)$ is
\be
B(k_1,k_2,k_3) = \frac{6}{5}f_{\rm NL} \left\{P_\zeta(k_1)P_\zeta(k_2) + {\rm perm.}\right\}\,.
\label{15}
\ee
This is of the so-called {\it local} form~\cite{shape}.
Equations~\eqref{14} and~\eqref{15} are consistent with $\zeta(x)$ of the form 
$\zeta(x)=\zeta_g(x)+\frac{3}{5}f_{NL}\zeta_g^2(x)$, where $\zeta_g$ is Gaussian.
The correlation function is evaluated at $t_{\rm end}$ ignoring gravity. 

Thus the 3-point function is fully specified by $f_{\rm NL}$~\cite{komat}. This parameter receives two contributions. To begin with,
the non-Gaussianity of $\delta\chi$, due to its cubic interaction in~(\ref{V2field}), is inherited by $\zeta$ through~(\ref{zetac}).
Following Maldacena~\cite{malda}, the $\delta\chi$ 3-point function is given by
\bea
\nonumber
\langle \delta\chi_1\delta\chi_2\delta\chi_3\rangle &=& -i\int_{-\infty}^{t_{\rm end}}ds\langle 0|[\delta\chi_1\delta\chi_2\delta\chi_3,{\cal H}_{\rm int}(s)]|0\rangle  \\
& & +\; {\rm c.\;c.}\,,
\eea
where $\delta\chi_i \equiv \delta\chi(x_i)$, and ${\cal H}_{\rm int}$ is the cubic interaction Hamiltonian from~(\ref{V2field}): ${\cal H}_{\rm int} = {\cal V}(\bar{\phi})\alpha_3\chi^3/3!\Lambda^3$.
An explicit calculation yields the {\it intrinsic} contribution
\be
f_{\rm NL}^{\rm int} = \mp\frac{5}{24}\frac{\alpha_3}{\beta\epsilon}\,.
\label{fint}
\ee
The $\mp$ sign corresponds to choosing 
$V_{,\chi}|$ to be $\pm$. 

The second contribution comes from the non-linear relation between $\delta\chi$ and $\zeta$ embodied in~(\ref{zetac}) and~(\ref{zetanonlin}). Even if $\delta\chi$ were
Gaussian, this non-linearity would make $\zeta$ non-Gaussian. This {\it conversion} contribution to $f_{\rm NL}$ is:
\be
f_{\rm NL}^{\rm conv} = \frac{5}{24}\frac{1}{\beta^2\epsilon}\,.
\label{fconv}
\ee

Summing~(\ref{fint}) and~(\ref{fconv}) yields a combined $f_{\rm NL}$:
\be
f_{\rm NL} \equiv  f_{\rm NL}^{\rm int} + f_{\rm NL}^{\rm conv} = \frac{5}{24\beta^2\epsilon}\left(1\mp\alpha_3\beta\right)\,.
\ee
Since this is inversely proportional to $\epsilon\ll 1$, non-Gaussianity tends to be large in New Ekpyrotic Cosmology.
Related ekpyrotic models~\cite{paolo,wands1} also give $f_{\rm NL}\sim \epsilon^{-1}$.
(A ghost condensate bounce and second scalar field are also invoked in~\cite{paolo}, albeit without an explicit
conversion mechanism; and while the two-field ekpyrotic phase of~\cite{wands1} is similar to ours, the bounce physics remains unspecified.)
This is in sharp contrast with slow-roll inflation, where $f_{\rm NL}$ is {\it proportional} to the slow-roll parameters and therefore unobservably small. 
For concreteness, consider our fiducial model with GUT-scale reheating, $\beta = 1$ and
$\epsilon = 10^{-2}$. 
Taking, for example, the $-$ sign in~\eqref{fint} and choosing
$2.728 > \alpha_3 >-3.8$ yields $f_{\rm NL}$ 
within the present WMAP 2$\sigma$ range: $-36< f_{\rm NL} < 100$. Thus $\alpha_3\sim {\cal O}(1)$ yields a non-Gaussian signal near the WMAP bound. Lower reheating temperatures correspond to smaller $\epsilon$ and, therefore, larger non-Gaussian signal. 
Of course, $|f_{\rm NL}|$ can always be made smaller by taking $\beta$, $\epsilon$ to 
be larger and/or by suitably choosing $\alpha_3$.

{\it 4-point function}: The connected 4-point function,
\bea
\nonumber
& & \langle \zeta_{k_1} \zeta_{k_2} \zeta_{k_3} \zeta_{k_4}\rangle = (2\pi)^3\delta^3(\vec{k}_1+\vec{k}_2+\vec{k}_3+\vec{k}_4) \\
& &\;\;\;\;\;\;\;\;\;\;\;\;\;\cdot [T(k_1,k_2,k_3,k_4) + T'(k_1,k_2,k_3,k_4)],
\label{4pt}
\eea
involves two different shape functions, evaluated at $t_{\rm end}$:
\bea
\nonumber
T &=& \frac{1}{2}\tau_{\rm NL} \left\{P_\zeta(k_1)P_\zeta(k_2)P_\zeta(k_{14}) + 23 \;{\rm perm.}\right\}\;; \\
T' &=& \kappa_{\rm NL}\left\{P_\zeta(k_1)P_\zeta(k_2)P_\zeta(k_3) + 3 \;{\rm perm.}\right\},
\label{4pt1}
\eea
where $\vec{k}_{ij} \equiv \vec{k}_i +\vec{k}_j$. Thus $T$ and $T'$ are specified respectively by the $\tau_{\rm NL}$ and $\kappa_{\rm NL}$ parameters. 
(Note that $\kappa_{\rm NL}$ is proportional to the $f_2$ parameter of~\cite{tauwmap}.) 
Equations~\eqref{4pt} and~\eqref{4pt1} are consistent with $\zeta(x)$ of the form 
$\zeta(x)=\zeta_g(x)+\frac{\sqrt{\tau_{\rm NL}}}{2}\zeta_g^2(x)+\frac{\kappa_{\rm NL}}{6}\zeta_g^3(x)$,
where $\zeta_g$ is Gaussian.
%Again, $\tau_{\rm NL}$ and $\kappa_{\rm NL}$ receive contributions 
%due to cubic and quartic $\delta\chi$ interactions as well 
%from the non-linear relation between $\delta\chi$ and $\zeta$.
Note that we can obtain $\tau_{\rm NL}$ immediately by simply comparing its definition with that of 
$f_{\rm NL}$: 
\begin{equation}
\tau_{\rm NL}=\frac{36}{25}f_{\rm NL}^2=\frac{1}{16\beta^4 \epsilon^2}
\left(1 \mp \alpha_3 \beta\right)^2\,. 
\label{1}
\end{equation}
This was also checked by explicitly computing the three and four-point functions.

Let us now consider $\kappa_{\rm NL}$.
It receives  two contributions: 
i) an intrinsic piece due to cubic and quartic $\delta\chi$ interactions; 
ii)  a conversion piece from the non-linear relation between $\delta\chi$ and $\zeta$.
The first contribution arises from cubic and quartic terms in $\chi$ in the 
potential~(\ref{V2field}). An explicit calculation gives
\be
\kappa_{\rm NL}^{\rm int} = \frac{2\alpha_4+3\alpha_3^2}{40\beta^2\epsilon^2}\,.
\ee
The second contribution is encoded in the $\zeta_c^2$ and $\zeta_c^3$ terms  in~(\ref{zetanonlin}). 
Comparing with~(\ref{4pt}), we obtain
\be
\kappa_{\rm NL}^{\rm conv} = \mp\frac{5\alpha_3}{24\beta^3\epsilon^2}\,.
\ee
Combining the above results, we find
\be
\kappa_{\rm NL} \equiv  \kappa_{\rm NL}^{\rm int} + \kappa_{\rm NL}^{\rm conv}  = 
\frac{\alpha_3(9\alpha_3\beta\mp 25) + 6\alpha_4\beta}{120\beta^3\epsilon^2}\,.
\ee

Both $\tau_{\rm NL}$ and $\kappa_{\rm NL}$
are inversely proportional to $\epsilon^2$ and therefore also tend to be relatively large. 
Note that $\tau_{\rm NL}$ is always positive, whereas $\kappa_{\rm NL}$ can be positive, 
zero or negative depending on the choices of $\alpha_{3}$ and $\alpha_{4}$. 
For instance, our fiducial parameter values for GUT-scale reheating with 
$\alpha_3,\alpha_4\sim {\cal O}(1)$ yield $\tau_{\rm NL}\sim 10^4$, which is around 
the estimated bound for the WMAP experiment~\cite{tauwmap}. 
Lower non-Gaussianity can again be achieved by taking larger $\beta$, 
$\epsilon$ and/or by a suitable choice of $\alpha_3$ and $\alpha_4$.

%\section{Discussion}

{\it Discussion}:
The simplest inflationary models, consisting of one or more slowly-rolling scalar fields, all predict negligible 3-point and higher-order correlation functions. Non-Gaussianity therefore offers a robust test to distinguish New Ekpyrotic Cosmology from slow-roll inflation. 

Significant inflationary non-Gaussianity can be obtained in non-slow-roll models, such as DBI inflation, albeit with a distinguishable shape dependence.
Our 3-point function is ``local", characterized by a momentum dependence that peaks for squeezed triangles, whereas the DBI amplitude peaks for 
equilateral triangles~\cite{shape}.

Large non-Gaussianity may also be achieved in the curvaton scenario. 
%although at the price of significant fine-tuning. 
%In that model, $f_{\rm NL}\sim 1/\Omega_{\rm dec}$, where $\Omega_{\rm dec}$ 
%is the fractional energy density in the curvaton when
%it decays~\cite{curvaton,lythng}. But since $\Omega_{\rm dec}$ fluctuates from 
%patch to patch, $f_{\rm NL}$ is generically either ${\cal O}(1)$ or unacceptably 
%large nearly everywhere. 
The curvaton 3-point function is also of the local form 
and hence cannot be used to distinguish curvatons from New Ekpyrotic Cosmology.
There is, however, an essential difference at the 4-point level. In the simplest curvaton model, the progenitor of density perturbations is a free field.
Thus, $\kappa_{\rm NL} \sim f_{\rm NL}$~\cite{lythng}. In contrast, in New Ekpyrosis, $\tau_{\rm NL}$ and $\kappa_{\rm NL}$ are generically of comparable
magnitude ($\sim \epsilon^{-2}$) and are expected to exhibit a distinguishable shape dependence. More intricate curvaton models with self-interactions
can also yield large $\kappa_{\rm NL}$. Similarly for general modulon scenarios~\cite{zal}.

Near-future non-Gaussianity observations will, therefore, test the new ekpyrotic paradigm and can potentially distinguish it from its inflationary alternatives.

In this paper we have used a simplifying approximation so as to obtain an analytic expression for non-Gaussianity.  
Our results give the exact parametric dependence while the details of the 
potential, roll-off time and so on are encoded in model 
dependent parameters, such as $\beta$.
We have checked that different quasi-analytic approximations continue to 
give the same parametric dependence as presented here, although with model-dependent coefficients that
can at most differ from those in this paper by factors of order unity.

{\it Acknowledgments:} We thank E.~Komatsu, P.J.~Steinhardt, A.~Tolley, and especially F.~Vernizzi for helpful discussions. 
This work is supported in part by NSERC and MRI (E.B. and J.K.), and by the DOE under
contract No. DE-AC02-76-ER-03071 and the NSF Focused Research Grant DMS0139799 (B.A.O.).


\begin{thebibliography}{hello} 

\bibitem{oldek}
 J.~Khoury, B.~A.~Ovrut, P.~J.~Steinhardt and N.~Turok,
%  ``The ekpyrotic universe: Colliding branes and the origin of the hot big bang,''
  Phys.\ Rev.\  D {\bf 64}, 123522 (2001); arXiv:hep-th/0105212;
  Phys.\ Rev.\  D {\bf 66}, 046005 (2002);
  J.~Khoury, B.~A.~Ovrut, N.~Seiberg, P.~J.~Steinhardt and N.~Turok,
  %``From big crunch to big bang,''
  Phys.\ Rev.\  D {\bf 65}, 086007 (2002).  
  
\bibitem{tolley}
  A.~J.~Tolley, N.~Turok and P.~J.~Steinhardt,
%  ``Cosmological perturbations in a big crunch / big bang space-time,''
  Phys.\ Rev.\ D {\bf 69}, 106005 (2004);   
    P.~L.~McFadden, N.~Turok and P.~J.~Steinhardt,
  %``Solution of a braneworld big crunch / big bang cosmology,''
  Phys.\ Rev.\  D {\bf 76}, 104038 (2007);
  T.~J.~Battefeld, S.~P.~Patil and R.~H.~Brandenberger,
  %``On the transfer of metric fluctuations when extra dimensions bounce or
  %stabilize,''
  Phys.\ Rev.\  D {\bf 73}, 086002 (2006).

\bibitem{newek}
  E.~I.~Buchbinder, J.~Khoury and B.~A.~Ovrut,
  %``New Ekpyrotic Cosmology,'' 
   Phys.\ Rev.\  D {\bf 76}, 123503 (2007);   
  %``On the Initial Conditions in New Ekpyrotic Cosmology,'' 
  JHEP {\bf 0711}, 076 (2007).
  
\bibitem{ghost}
 N.~Arkani-Hamed, H.C.~Cheng, M.A.~Luty and S.~Mukohyama,
  %``Ghost condensation and a consistent infrared modification of gravity,''
  JHEP {\bf 0405}, 074 (2004);
   P.~Creminelli {\it et al.},
%   P.~Creminelli, M.A.~Luty, A.~Nicolis and L.~Senatore,
  %``Starting the universe: Stable violation of the null energy condition and
  %non-standard cosmologies,''
  JHEP {\bf 0612}, 080 (2006).
  
\bibitem{lehners}
  F.~Finelli,
%  ``Assisted contraction,''
  Phys.\ Lett.\  B {\bf 545}, 1 (2002);
 J.~L.~Lehners {\it et al.},
% J.~L.~Lehners, P.~McFadden, N.~Turok and P.~J.~Steinhardt,
%``Generating ekpyrotic curvature perturbations before the big bang,''
  Phys.\ Rev.\  D {\bf 76}, 103501 (2007).
  
\bibitem{paolo}
  P.~Creminelli and L.~Senatore,
%  ``A smooth bouncing cosmology with scale-invariant spectrum,''
  JCAP {\bf 0711} (2007) 010.    

\bibitem{WMAP}
  D.~N.~Spergel {\it et al.},
  %``Wilkinson Microwave Anisotropy Probe (WMAP) three year results:
  %Implications for cosmology,''
  Astrophys.\ J.\ Suppl.\  {\bf 170}, 377 (2007);
  P.~Creminelli {\it et al.},
%  P.~Creminelli, L.~Senatore, M.~Zaldarriaga and M.~Tegmark,
  %``Limits on f_NL parameters from WMAP 3yr data,''
  JCAP {\bf 0703}, 005 (2007).
  
\bibitem{tauwmap}
  N.~Kogo and E.~Komatsu,
  %``Angular Trispectrum of CMB Temperature Anisotropy from Primordial
  %Non-Gaussianity with the Full Radiation Transfer Function,''
  Phys.\ Rev.\  D {\bf 73}, 083007 (2006).
  
\bibitem{DBI}
  M.~Alishahiha, E.~Silverstein and D.~Tong,
  %``DBI in the sky,''
  Phys.\ Rev.\  D {\bf 70}, 123505 (2004).
    

\bibitem{curvaton}
  D.~H.~Lyth and D.~Wands,
  %``Generating the curvature perturbation without an inflaton,''
  Phys.\ Lett.\  B {\bf 524}, 5 (2002);
    K.~Enqvist and M.~S.~Sloth,
  %``Adiabatic CMB perturbations in pre big bang string cosmology,''
  Nucl.\ Phys.\  B {\bf 626}, 395 (2002);
    T.~Moroi and T.~Takahashi,
  %``Effects of cosmological moduli fields on cosmic microwave background,''
  Phys.\ Lett.\  B {\bf 522}, 215 (2001); [Erratum-ibid.\  B {\bf 539}, 303 (2002)].

  \bibitem{lythng}
  D.~H.~Lyth,
  %``Non-gaussianity and cosmic uncertainty in curvaton-type models,''
  JCAP {\bf 0606}, 015 (2006); 
   M.~Sasaki, J.~Valiviita and D.~Wands, 
   %``Non-gaussianity of the primordial perturbation in the curvaton model,''
   Phys.\ Rev.\ D {\bf 74}, 103003 (2006). 
%  F.~Bernardeau and J.~P.~Uzan,
%  %``Inflationary models inducing non-gaussian metric fluctuations,''
%  Phys.\ Rev.\  D {\bf 67}, 121301 (2003).

\bibitem{dgz}
  G.~Dvali, A.~Gruzinov and M.~Zaldarriaga,
  %``A new mechanism for generating density perturbations from inflation,''
  Phys.\ Rev.\  D {\bf 69}, 023505 (2004); 
    L.~Kofman,
  %``Probing string theory with modulated cosmological fluctuations,''
  arXiv:astro-ph/0303614.

\bibitem{zal}
  M.~Zaldarriaga,
  %``Non-Gaussianities in models with a varying inflaton decay rate,''
  Phys.\ Rev.\  D {\bf 69}, 043508 (2004);
    F.~Vernizzi,
  %``Cosmological perturbations from varying masses and couplings,''
  Phys.\ Rev.\  D {\bf 69}, 083526 (2004).

\bibitem{gradient}
  D.~S.~Salopek and J.~R.~Bond,
  %``Nonlinear evolution of long wavelength metric fluctuations in inflationary
  %models,''
  Phys.\ Rev.\  D {\bf 42}, 3936 (1990);
  D.~H.~Lyth, K.~A.~Malik and M.~Sasaki,
  %``A general proof of the conservation of the curvature perturbation,''
  JCAP {\bf 0505}, 004 (2005).

\bibitem{shape}
 D.~Babich, P.~Creminelli and M.~Zaldarriaga,
  %``The shape of non-Gaussianities,''
  JCAP {\bf 0408}, 009 (2004).

\bibitem{komat}
  E.~Komatsu and D.~N.~Spergel,
  %``Acoustic signatures in the primary microwave background bispectrum,''
  Phys.\ Rev.\  D {\bf 63}, 063002 (2001).

\bibitem{wands1}
 K.~Koyama and D.~Wands,
%``Ekpyrotic collapse with multiple fields,''
JCAP {\bf 0704}, 008 (2007);
    K.~Koyama {\it et al.},
%    K.~Koyama, S.~Mizuno, F.~Vernizzi and D.~Wands,
  %``Non-Gaussianities from ekpyrotic collapse with multiple fields,''
  JCAP {\bf 0711}, 024 (2007).

\bibitem{malda}
  J.~M.~Maldacena,
%  ``Non-Gaussian features of primordial fluctuations in single field
%  inflationary models,''
  JHEP {\bf 0305}, 013 (2003).






\end{thebibliography}
\end{document}